# An effective approach for classification of advanced malware with high accuracy

Ashu Sharma[1] and Sanjay K. Sahay[2,*]

*[1,2]Department of Computer Science and Information System,*
*[1,2]Birla Institute of Technology and Science, K. K. Birla Goa Campus, NH-17B, By Pass Road, Zuarinagar-403726, Goa, India*
*{[1]p2012011, [2]ssahay}@goa.bits-pilani.ac.in*

## *Abstract*

*Combating malware is very important for software/systems security, but to prevent the software/systems from the advanced malware, viz. metamorphic malware is a challenging task, as it changes the structure/code after each infection. Therefore in this paper, we present a novel approach to detect the advanced malware with high accuracy by analyzing the occurrence of opcodes (features) by grouping the executables. These groups are made on the basis of our earlier studies [1] that the difference between the sizes of any two malware generated by popular advanced malware kits viz. PS-MPC, G2 and NGVCK are within 5 KB. On the basis of obtained promising features, we studied the performance of thirteen classifiers using N-fold cross-validation available in machine learning tool WEKA. Among these thirteen classifiers we studied in-depth top five classifiers (Random forest, LMT, NBT, J48 and FT) and obtain more than 96.28% accuracy for the detection of unknown malware, which is better than the maximum detection accuracy (~95.9%) reported by Santos et al (2013). In these top five classifiers, our approach obtained a detection accuracy of ~97.95% by the Random forest.*

## 1. Introduction

"Malware refers to a program that is inserted into a system, usually covertly, with the intent of compromising the confidentiality, integrity, or availability of the victim's data, applications, or operating system or of otherwise annoying or disrupting the victim" [2]. From the last four decades, malware is continuously evolving with high complexity to evade the available detection technique. These malware are basically classified as first and second generation malware. In the first generation, structure of the malware does not change, while in the second generation, structure changes to generate a new variant, keeping the action same [3]. On the basis of how variances are created in malware, second generation malware are further classified into Encrypted, Oligormorphic, Polymorphic and Metamorphic Malware [4]. These malware changes its structure in random and unpredictable ways each time it replicates, hence hard to detect. According to McAfee technical report of 2014, there are more than 200 million known malware samples [5]. The Symantec 2014 Internet Security Threat report states that 2013 was the mega breach year [6] (~62% more breaches then 2012). The F-secure document reported an increase in malware attacks against mobile devices based on Android and Apple iOS [7]. This increase in threat from malware is due to wide spread use of World Wide Web. An estimate shows that the web-based attacks were increased 36% with over 4,500 new attacks each day, annoying/disrupting the victim in terms of confidentiality, integrity, availability of the victims data etc. [8]. The malware attacks/threat are not only limited to individual level, but there are state sponsored highly skilled hackers developing customized malware to disrupt industries and for military espionage [9]. Such attacks can alter the operation of industrial systems, disrupt power plants, e.g. the StuxNet and Duqu malware [10]. The intrusion into Google's systems demonstrates how well-organized attacks are designed to maintain long-term access to an organization's network [11].

* *ssahay*@goa.bits-pilani.ac.in

To combat threats/attacks from the malware, signature-based software (anti-malware) were widely deployed. However, its an indisputable fact that these traditional approach of combating the threats/attack with a technology-centric are ineffective to detect today's highly sophisticated customized malware. Hence attacks from such malware to the computing world are increasing every day. The consequence will be devastating if in time adequate measures had not been taken. Therefore, there is a need that both academia and anti-malware developers should continually work to combat the threats/attacks from the evolving malware. The most popular techniques used for the detection of malware are signature matching, heuristics-based detection, malware normalization, machine learning, etc. [4]. In recent years, machine learning techniques are studied by many authors and proposed different approaches [12] [13] [14], which can supplement traditional anti-malware system. For the detection of malware by machine learning technique, feature selection plays a vital role. In the literature, many feature selection approaches are discussed viz. Olivier Henchiri et al. 2006 [15], Siddiqui et al. 2008 [16], B. Mehdi et al. 2009 [17] and Santos et al. 2013 [18] used hierarchical, unary variable removal method, Goodness evaluator and Weighted Term Frequency (WTF) respectively for the feature selection. The maximum accuracy they obtained was 95.26%. In this paper, our approach outperforms the accuracy obtained by these authors by more than ～2%.

The paper is organized as follow, in next section related work is discussed and in section 3 we present our approach, The section 4 discuss the experimental results and finally section 5 contains the conclusion of the paper.

## 2. Related work

The first virus was created in 1970 [19] and since then there is a strong contest between the attackers and defenders. This rat-race led to the improvement in both malware and its detection techniques. To defend the malware attacks, anti-malware groups are developing new techniques. On the other hand, malware developers are adopting new tactics/methods to avoid the malware detectors. Initially, the tools and techniques of malware analysis were in the domain of anti-malware vendors. However, the use of malware for espionage, sophisticated cyber attacks and other crimes motivated the academicians and digital investigators to develop an advanced method to combat the threats/attacks from it. In the year 2001, Schultz et al. [20] were the first to introduce the concept of data mining for detecting malware. They used three different static features for malware classification (Portable Executable, strings and byte sequences). Kolter and Maloof (2004) evaluated data sets using Instance-Based Learning Algorithms (IBK), TF-IDF, Naive Bayes, Support Vector Machine (SVM) and Decision tree [21]. Among the classifiers used by them, Decision tree outperformed. In the year 2005, Karim et al. [22] addressed the tracking of malware evolution based on opcode sequences and permutations. O. Henchiri et al. (2006) proposed a hierarchical feature extraction algorithm and used ID3, j48, Naive Bayes and SMO classifier and obtained the maximum of 92.56% accuracy [15]. In the year 2007, Bilar uses small dataset (67 malware and 20 benign samples) to examine the difference in opcode frequency distribution between malicious and benign programs [23]. He found that malware opcode distribution differs significantly from benign programs and also observed that some opcodes seen to be a stronger predictor of frequency variation. He however, did not apply it for the classification of advanced malware. In the year 2008, Tian et al. particularly classified Trojan malware using function length frequency [24]. Their results indicate that the function length along with its frequency is significant in identifying malware family and can be combined with other features for fast and scalable malware classification. In the year 2008, Siddiqui et al. [16] used variable length instruction sequence for detecting worms in the wild. They tested their method on a data set of 2774 (1444 worms and 1330 benign files) and got 95.6% detection accuracy. In the year 2008, Moskovitch et al. [25] [26] compared the different classifiers by byte-sequence n-grams (3, 4, 5 or 6). Among the classifiers they studied BDT, DT and ANN outperformed NB, BNB and SVM classifiers, exhibiting lower false positive rates. S. Momina Tabish (2009) used AdaBoostM1 algorithm for classification by taking n-gram frequency as a feature and reported 90% detection accuracy [27]. In the year 2010, Bilal Mehdi et. al. [28] used hyper-grams (generalized n-gram) and obtained 87.85% detection accuracy and claimed no false alarm. Santos et al. in the year 2011 pointed out that supervised learning requires a significant amount of labeled executables for both malware and benign programs, which is difficult to obtain, hence they proposed a semi-supervised learning method for detecting unknown malware, which does not require a large amount of labeled data [29]. They obtained 86% of accuracy by labeling only 50% of the selected data set. In the subsequent paper [18] in 2013, they used Term Frequency for modeling different classifiers and found that SVM outperforms with accuracy of 92.92% and 95.90% respectively for one opcode and two opcode sequence length respectively.

Recently in 2014, Kevin Allix et al. [12] took a size-able dataset of over 50,000 android applications and implemented using 4 well-known machine learning algorithms (RandomForest, J48, JRip and LibSVM) with ten-fold cross-validation. He claimed his approach outperformed existing machine learning approaches, however on usual size datasets performance does not translate in the wild.

## 3. Our approach

In order to uncover the unknown malware with high accuracy, our novel approach as shown in Figure 1 involves finding the promising features (Algo. 1), training of classifiers and detection of unknown malware.

### 3.1. Building the Datasets and Feature Selection

To build the datasets, we downloaded 11088 malware from malacia-project [30] and collected 4006 benign programs (cross checked from virustotal.com [31]) from different systems.

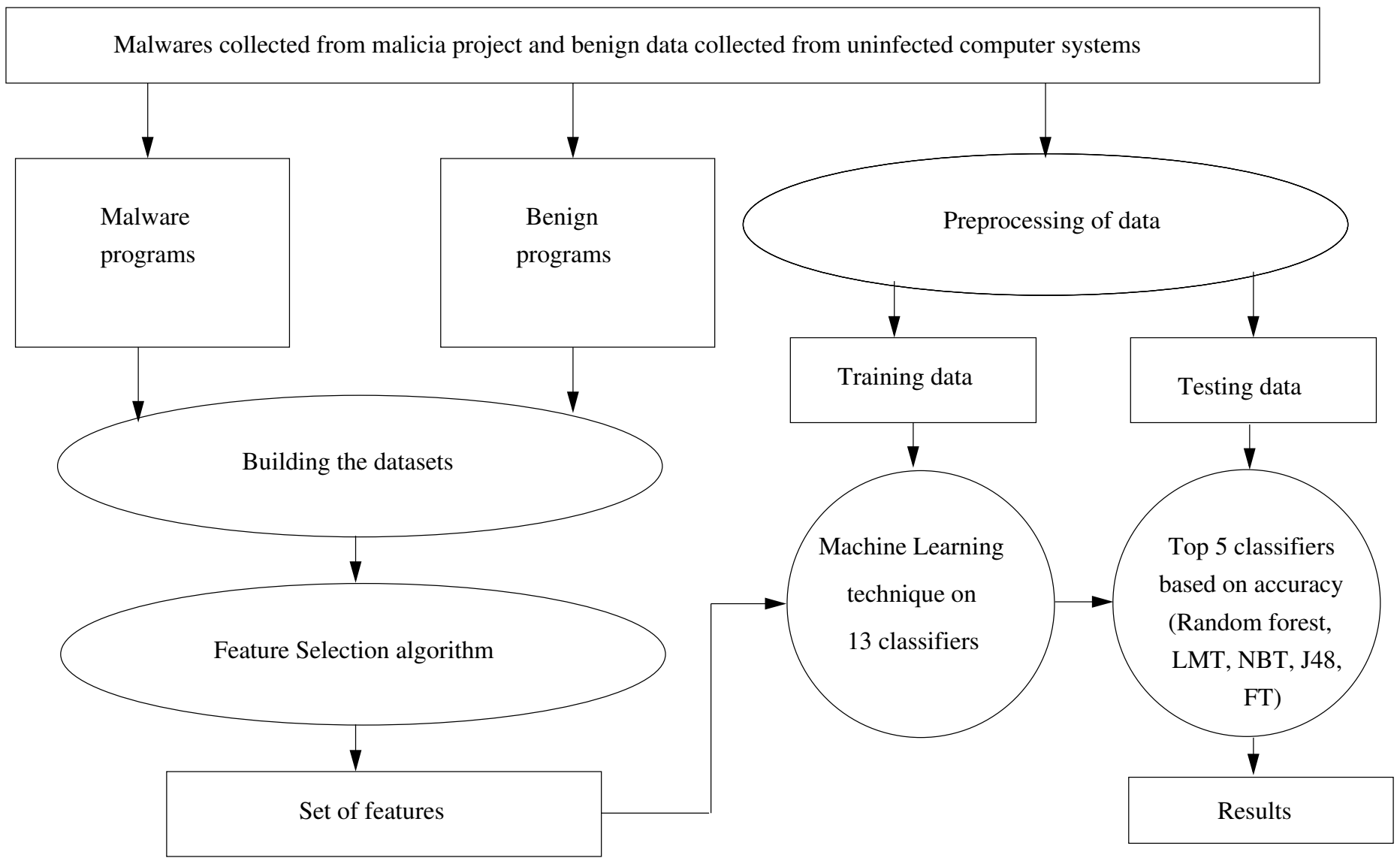

**Figure 1.** ***Flow chart for the detection of unknown malware.***

---

**Algorithm 1: Feature Selection**

---

**INPUT:** Pre-processed data of all groups,
$N_b$→ Number of benign executables and $N_m$→ Number of malware executables
**OUTPUT:** List of features.
**BEGIN**
**for all** groups $G_K$ **do**
  **for all** $i_{th}$ benign executable in the group $G_K$ **do**
    Compute the normalized frequency $F_{K_b}$ of each opcode $o_j$

$$F_{K_b}(o_j) = (\sum f_i(o_j)) / N_b$$

  **end for**
  **for all** $i_{th}$ malware executable in the group $G_K$ **do**
    Compute the normalized frequency $F_{K_m}$ of each opcode $o_j$

$$F_{K_m}(o_j) = (\sum f_i(o_j)) / N_m$$

  **end for**
  **for all** opcode $o_j$ **do**
    Subtract the frequencies $F_{k_b}$ and $F_{k_m}$.

$$D_K(o_j) = \left| F_{K_b}(o_j) - F_{K_m}(o_j) \right|$$

  **end for**
  Sort the obtained $D_K(o)$.
**end for**
Set a threshold on $D_k(o)$ to select the promising opcodes features such that from each $G_k$ at least 10 opcodes get selected.
**return** Union of the selected features of all the groups.

---

The promising features of the executables are obtained by clubbing the dataset in 5 KB size of 100 groups [1] as in the collected dataset ~97.18% malware are below 500 KB (Figure 2) and the difference between the sizes of any two malware generated by popular advanced malware kits viz. NGVCK [32], PS-MPC [33] and G2 [34] are within 5 KB. Hence, the features obtained will have a signature of maximum executables to detect the unknown malware. Our features are opcodes of the executables obtained by *objdump* utility available in the Linux system. To identify the each opcode we labeled it with a fixed integer. To differentiate malware and benign programs we obtained the features as given in algo. 1.

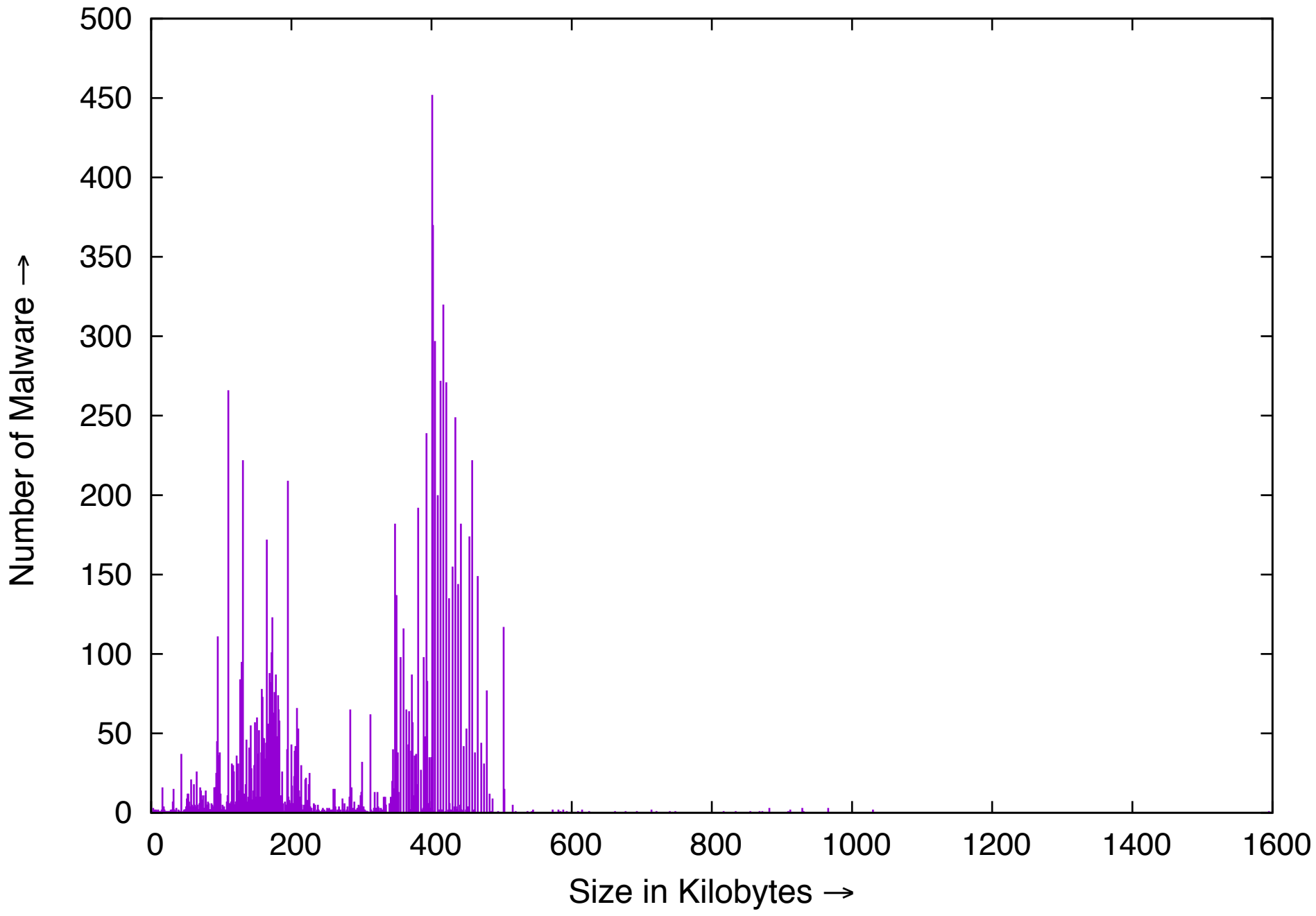

**Figure 2.** ***Number of malware with respect to file size.***

## 3.2. Training of the classifiers

To find the best classifiers for detection of unknown malware, we investigated thirteen tree based classifiers viz. Random forest, J48, REPTREE, LMT, Decision stump, ADT, NBT, FT, LAD, Random Tree, Simple CART, BFT and J48 Graft available in the popular and widely used suite of machine learning software known as *WEKA* (a collection of visualization tools and algorithms for data analysis and predictive modeling, together with graphical user interfaces for easy access to this functionality). Then with the obtained features, we run the *WEKA* n-fold cross-validation to train all the selected classifiers. Figure 3 shows the accuracy obtained by all classifiers for $n$ = 2,4,6...,16 folds. We observed that Random forest is the best classifier and its accuracy is almost flat after $n$ = 2. Rest twelve classifiers accuracy fluctuates, however after ten-fold cross-validation the fluctuations are least and we observe maximum correctness in the accuracy.

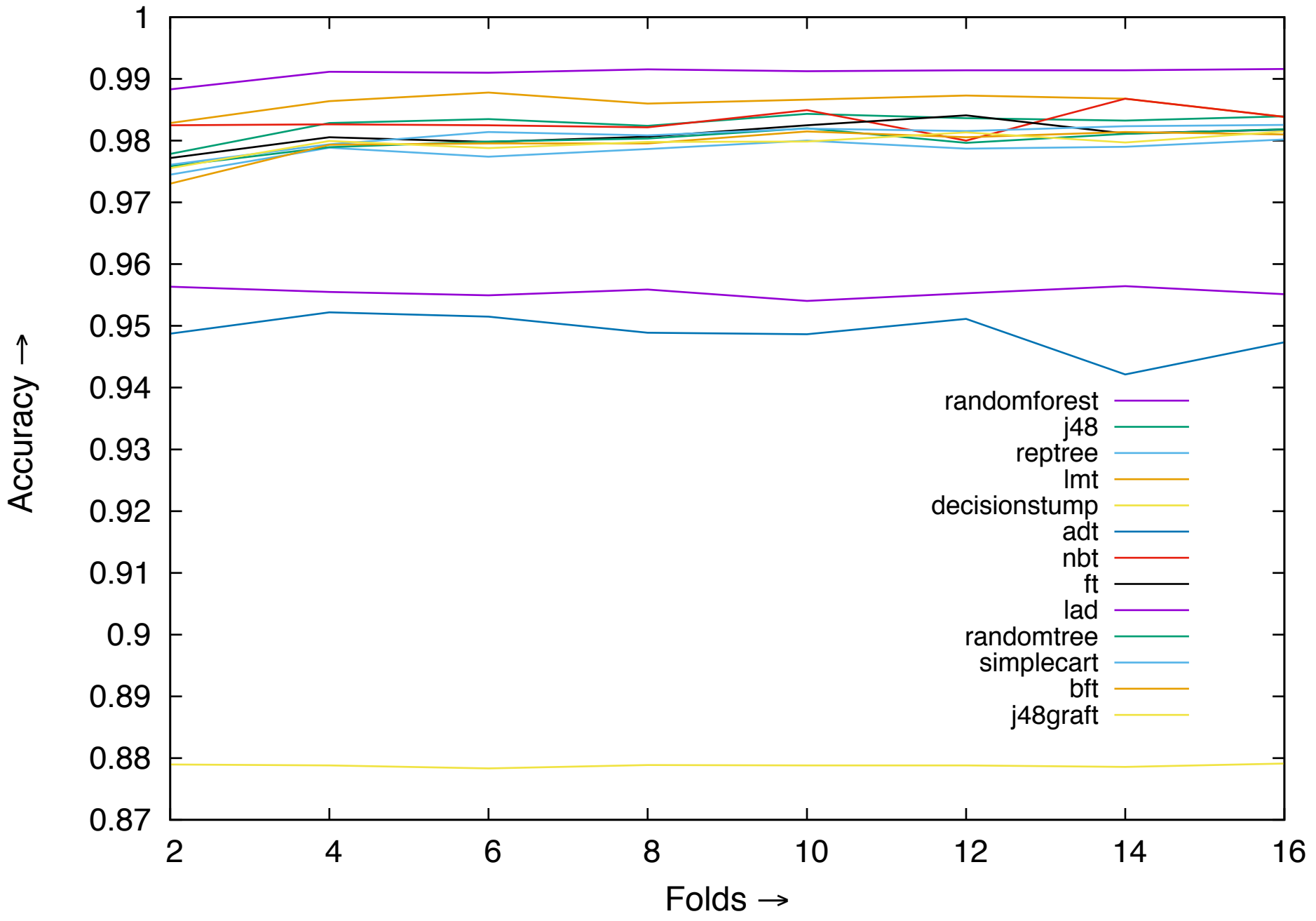

**Figure 3.** ***Accuracy of the thirteen classifiers with N-fold cross validation.***

## 3.3 Detection of unknown malware

Among the thirteen studied classifier, we selected top five (Random forest, LMT, NBT, J48 and FT) for in-depth analysis. To study the overall performance of these classifiers, we randomly selected 750 malware and 610 benign programs from all the groups, such that at least five executables from each group can be randomly tested by the trained classifiers with ten-fold cross-validation for the detection of unknown malware.

## 4. Experimental Results

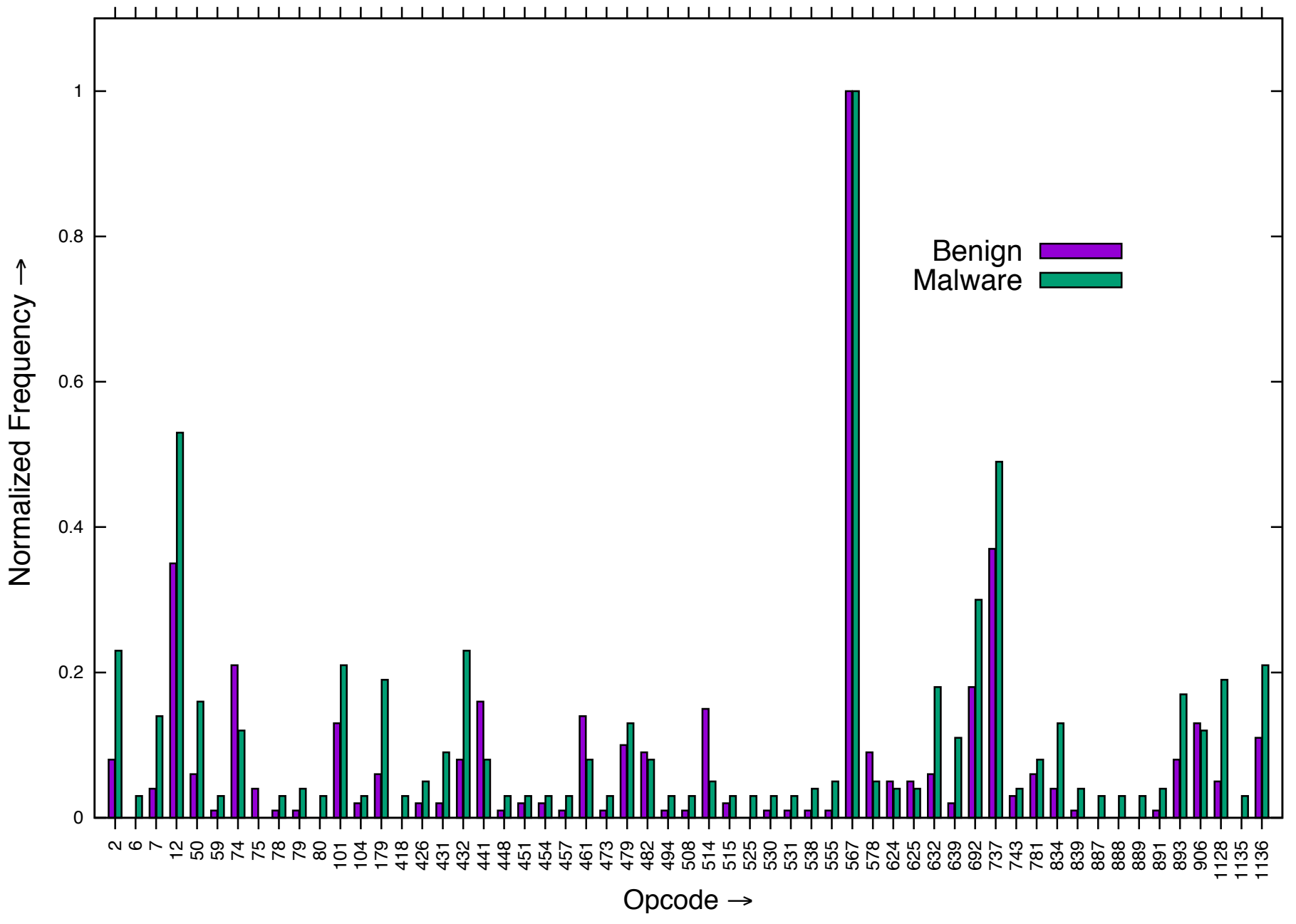

**Figure 4.** ***Normalized opcode occurrence of all the collected malware and benign program keeping threshold 0.02.***

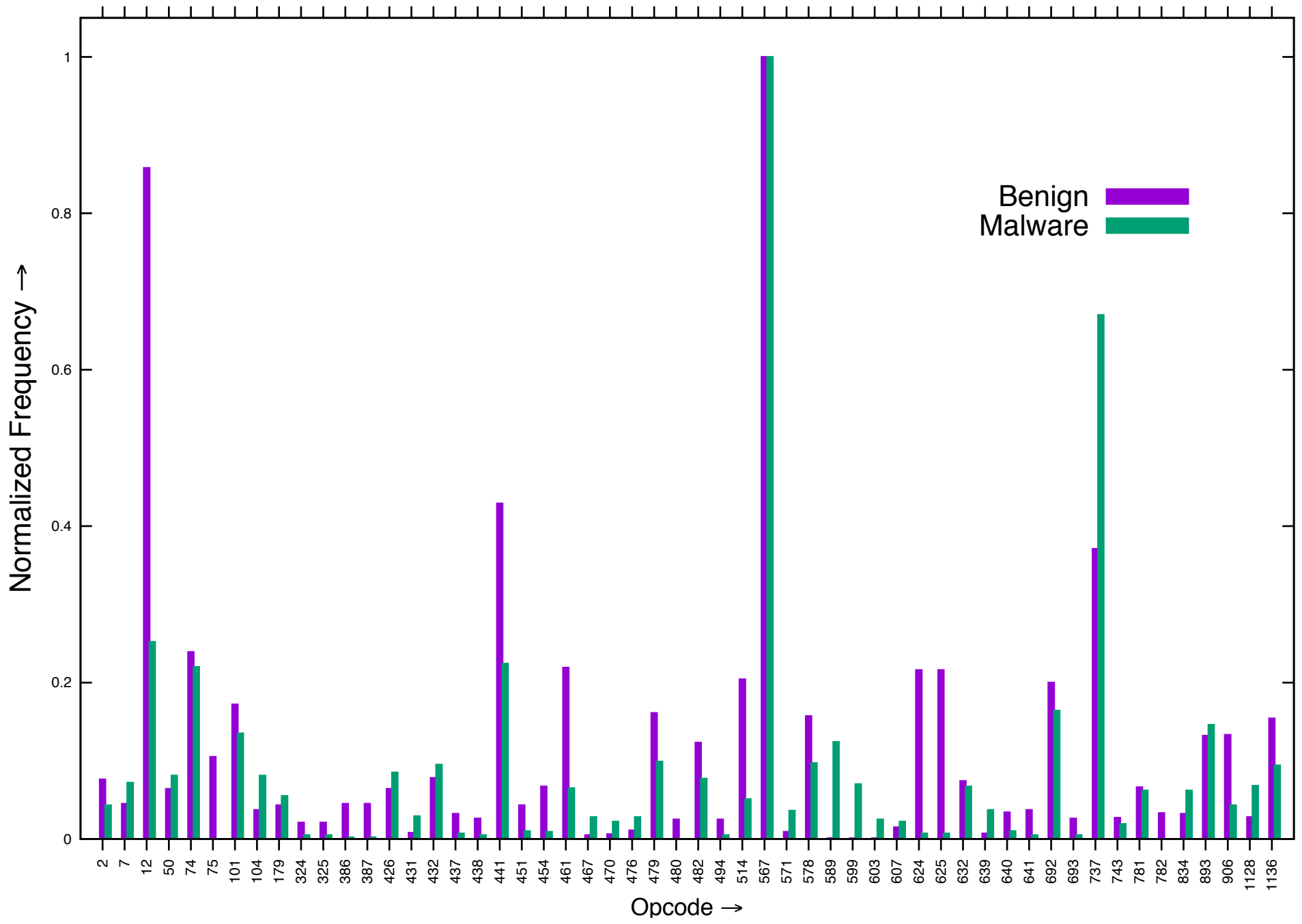

**Figure 5.** ***Normalized opcode occurrence of the malware and benign program of size 10-15 KB keeping threshold 0.02.***

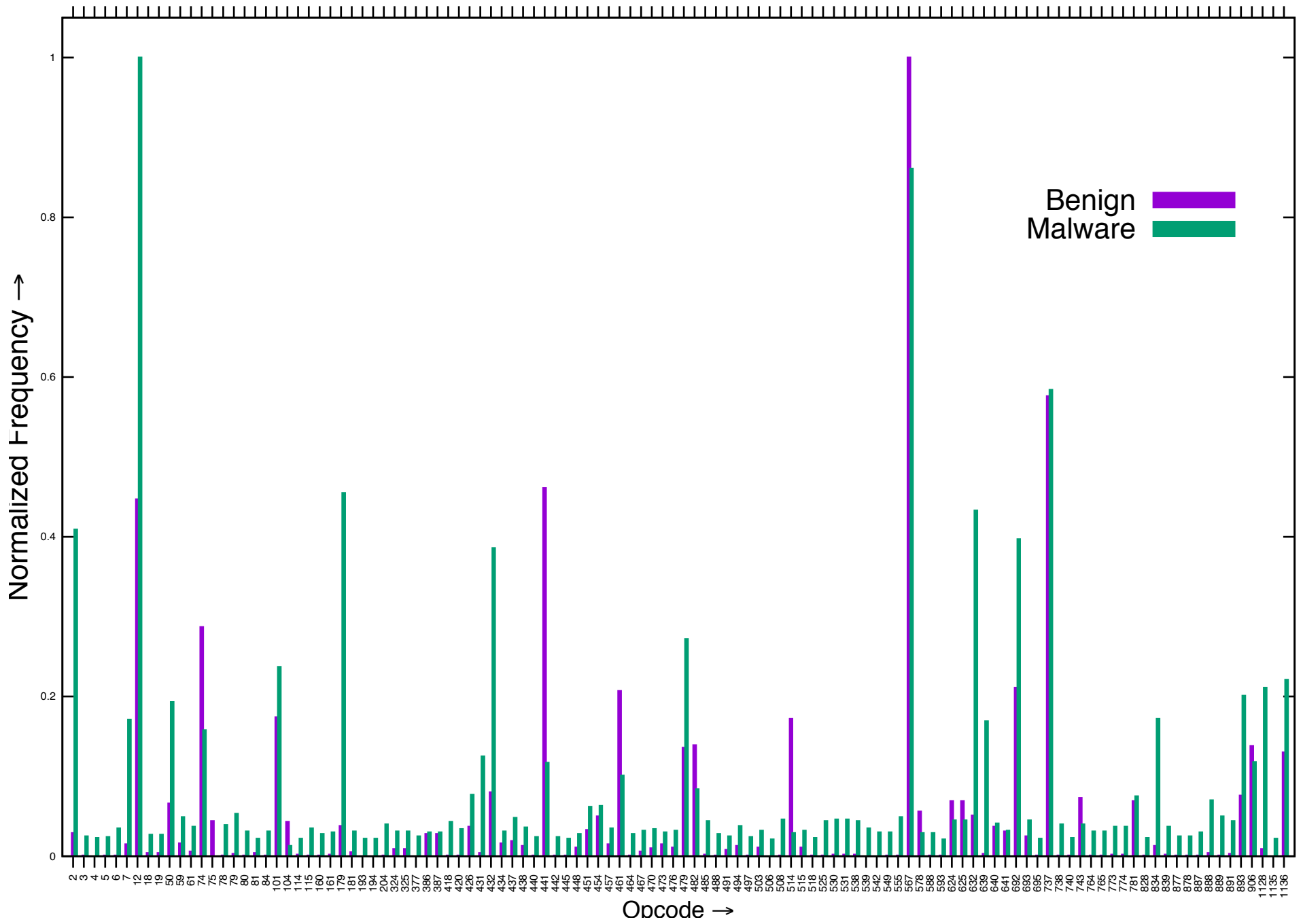

**Figure 6.** ***Normalized opcode occurrence of the malware and benign program of size 140-145 KB keeping threshold 0.02.***

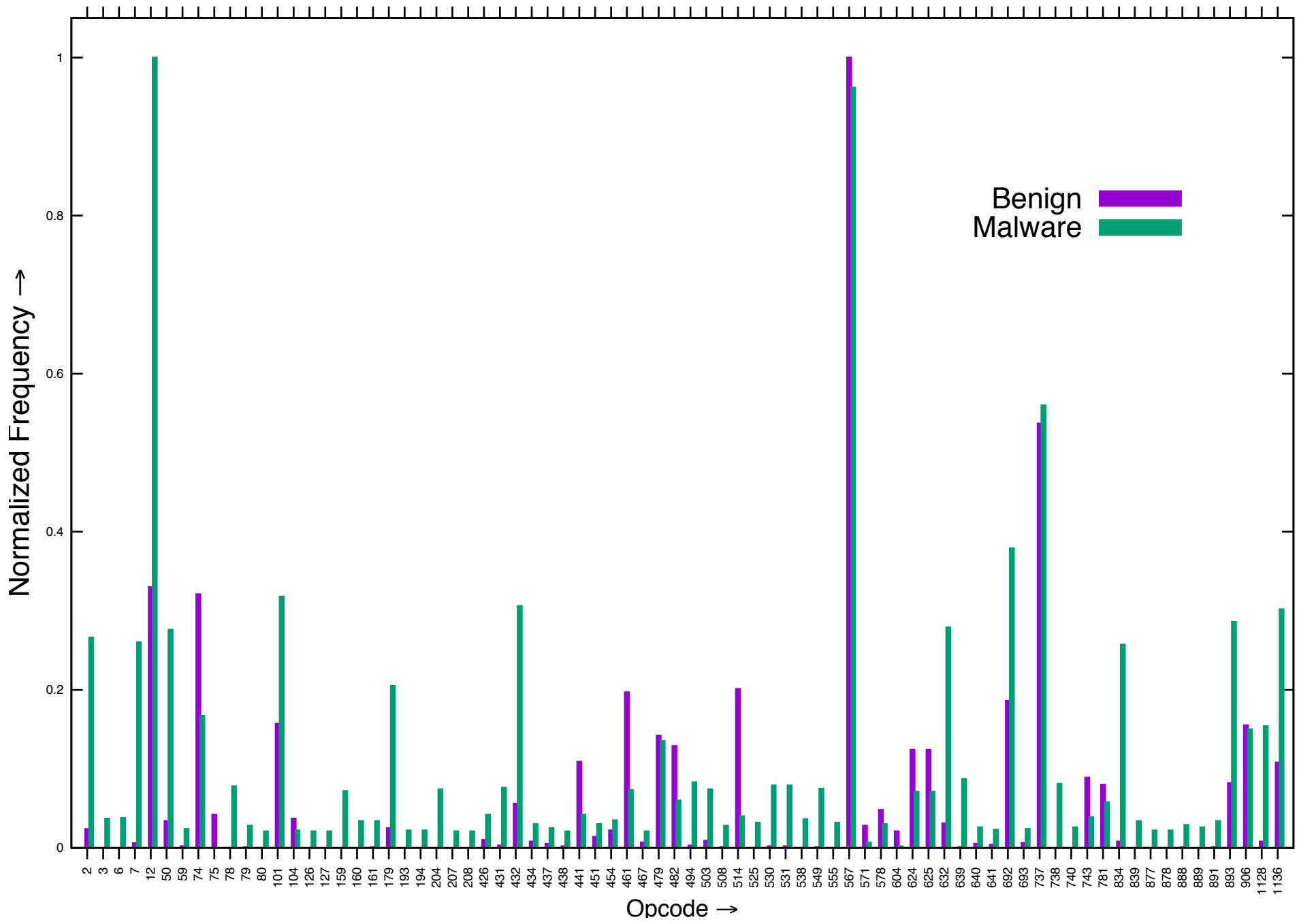

**Figure 7.** ***Normalized opcode occurrence of the malware and benign program of size 240-245 KB keeping threshold 0.02.***

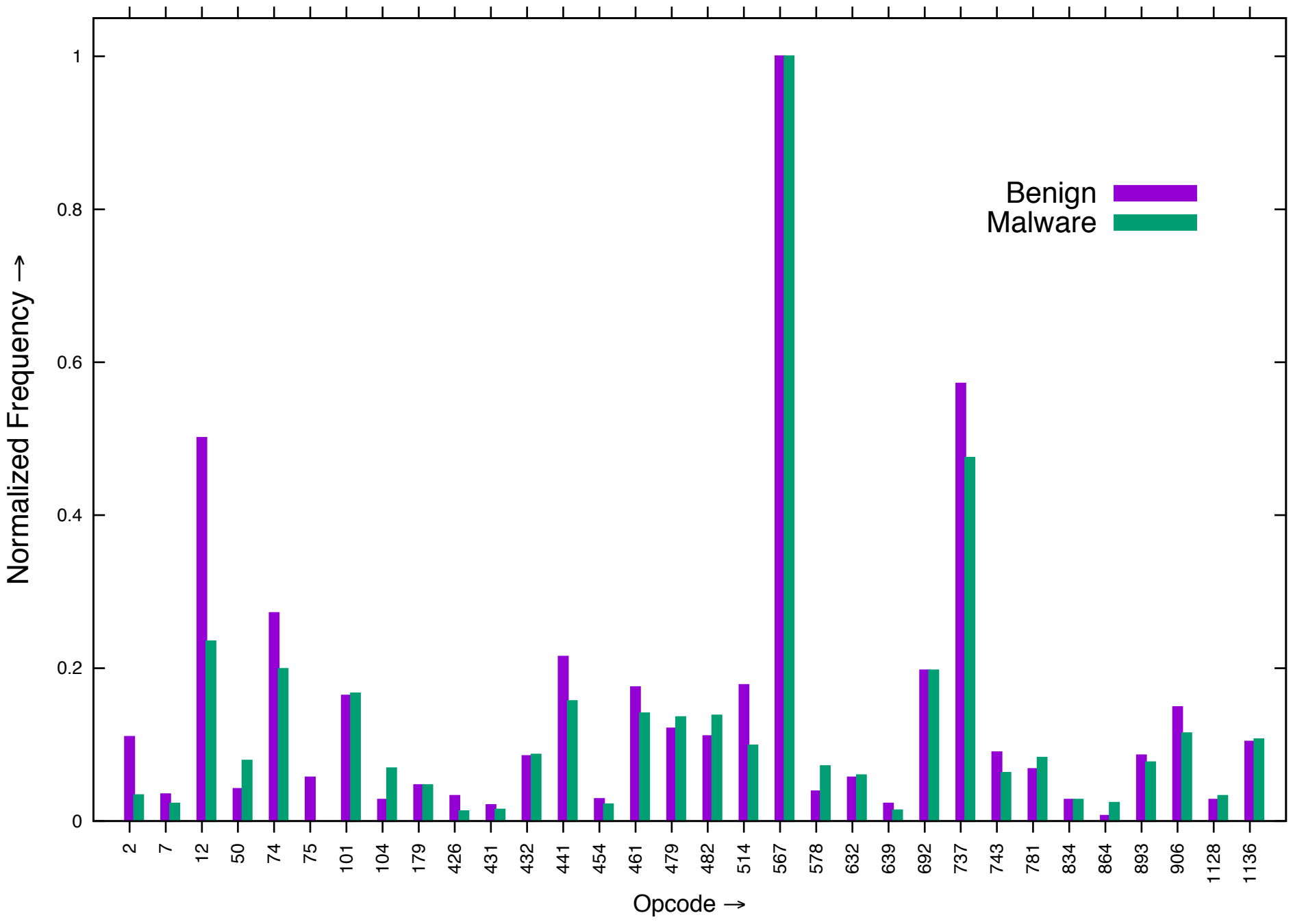

**Figure 8.** ***Normalized opcode occurrence of the malware and benign program of size 415-420 KB keeping threshold 0.02.***

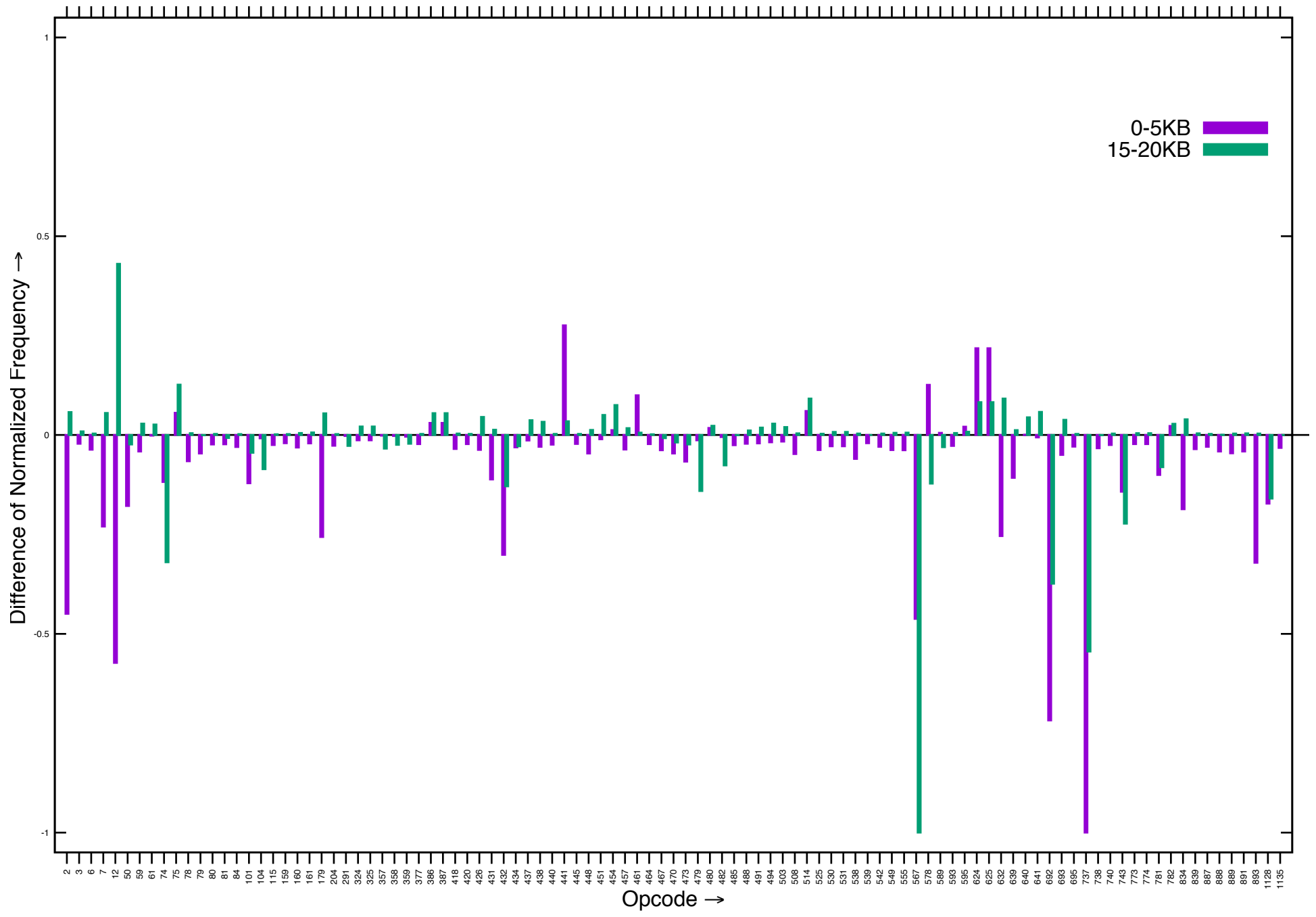

**Figure 9.** ***Difference in the occurrence of respective opcodes between benign and malware program of size 0-5 KB and 15-20 KB keeping threshold 0.02.***

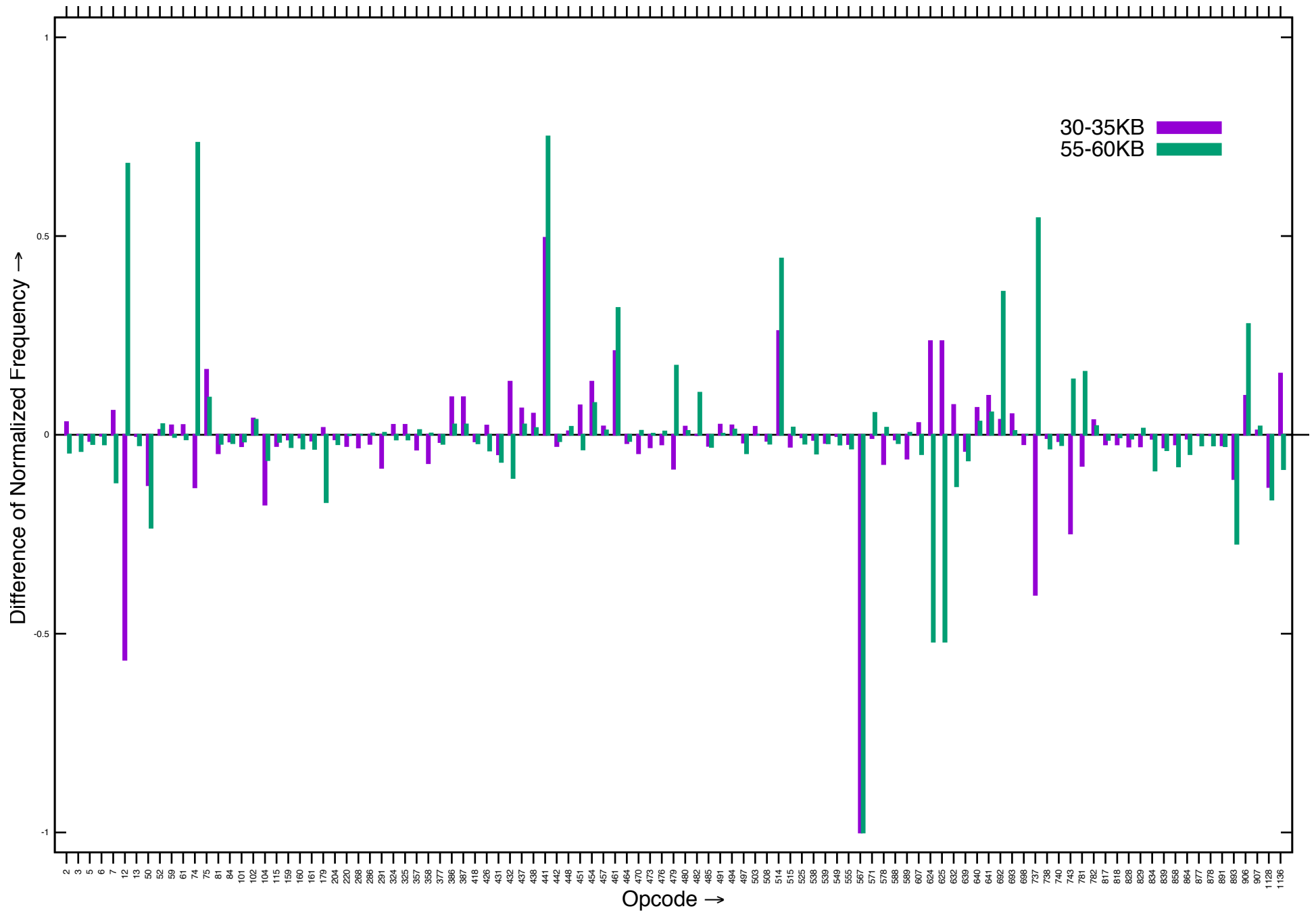

**Figure 10.** ***Difference in the occurrence of respective opcodes between benign and malware program of size 30-35 KB and 55-60 KB keeping threshold 0.02.***

As discussed in section 3, we disassembled all the malware and benign programs of below 500 KB size. Then we computed the normalized opcode occurrence of all malware and benign programs, for each group separately. We observed that the opcode occurrence in the malware and benign programs differ in large. Fig. 4 shows the normalized opcode occurrence of all the malware and benign programs and fig 5, 6, 7, & 8 shows the normalized opcode occurrence for the group 10-15, 140-145, 240-245 & 415-420 KB size respectively keeping the lower threshold 0.02. To find the dominant features of malware and benign programs we computed the difference in opcode occurrence between the benign and malware programs of each group and found that dominant opcodes vary from group to group. Figure 9. shows the difference in the occurrence of respective opcodes between benign and malware program of size 0-5 KB and 15-20 KB size keeping the lower threshold 0.02. Similarly, fig 10 is plotted for the size 30-35 and 55-60 KB size of the data set. In the figure 9, 10 upper side shows the opcodes that dominate in benign and lower side shows the opcode that dominates in malware.

The effectiveness of the top five classifiers viz. Random forest Tree, LMT, NBT, J48 and FT has been studied with the randomly selected 750 malware and 610 benign programs. The analysis are done in *WEKA* with ten-fold cross-validation, in terms of True Positive Ratio (TPR), True Negative Ratio (TNR), False Positive Ratio (FPR), False Negative Ratio (FNR) and accuracy, defined as

$$TPR=\frac{TP}{TM}; \quad TNR=\frac{TN}{TB}; \quad FPR=\frac{FP}{TB}; \quad FNR=\frac{FN}{TM}$$

$$Accuracy=\frac{TP+TN}{TM+TB}\times 100$$

where,
$TP\longrightarrow$ True positive, the number of malware correctly classified
$TN\longrightarrow$ True negative, the number of benign correctly classified.
$FP\longrightarrow$ False positive, the number of benign detected as malware.
$FN\longrightarrow$ False negative, the number of malware detected as benign.
$TM\longrightarrow$ Total number of malware.
$TB\longrightarrow$ Total number of benign.

From the analysis, it is clear that Random forest is the best classifier for identification of unknown malware. Nevertheless, the other classifiers are also reasonably good (> 96.2%) for the detection of unknown malware. The detail results obtained are shown in table 1.

**Table 1. Performance of the top 5 classifiers.**

| Classifiers | True positive | False negative | False positive | True Negative | Accuracy |
|---|---|---|---|---|---|
| Random forest | 739 (98.53%) | 11 (1.47%) | 6 (2.81%) | 554 (97.19%) | 97.95% |
| LMT | 734 (97.87%) | 16 (2.13%) | 23 (4.04%) | 547 (95.96%) | 97.04% |
| NBT | 728 (97.07%) | 22(2.93%) | 19 (3.33%) | 551 (96.67%) | 96.89% |
| J48 | 729 (97.2%) | 21 (2.8%) | 23 (4.04%) | 547 (95.96%) | 96.66% |
| FT | 729 (97.2%) | 21 (2.8%) | 8 (4.91%) | 542 (95.09%) | 96.28% |

We found that the classifier NBT, J48 and FT have the almost same True positive ratio. In these, the overall accuracy of Functional Tree classifier is lowest, which is basically due to high False positive ratio. Figure 11 shows the variation of False positive and False negative of the studied classifiers. We found that the False positives ratio of Random forest and LMT are almost double than False positive ratio, however for overall accuracy both has to be low. From Figure 12 we observed that the True positives of all classifiers are more biased towards the detection of malware. The obtained accuracy for all five classifiers are shown in Figure 13 and the comparison of our results with Santos et al., Siddiqui et al., Asaf Shabtai et. al. for Random forest and Mehdi et al., Santos et al., Olivier Henchiri et al. for J48 are shown in Figure 14. Among these authors our approach uncover the malware with the best accuracy (~97.95%).

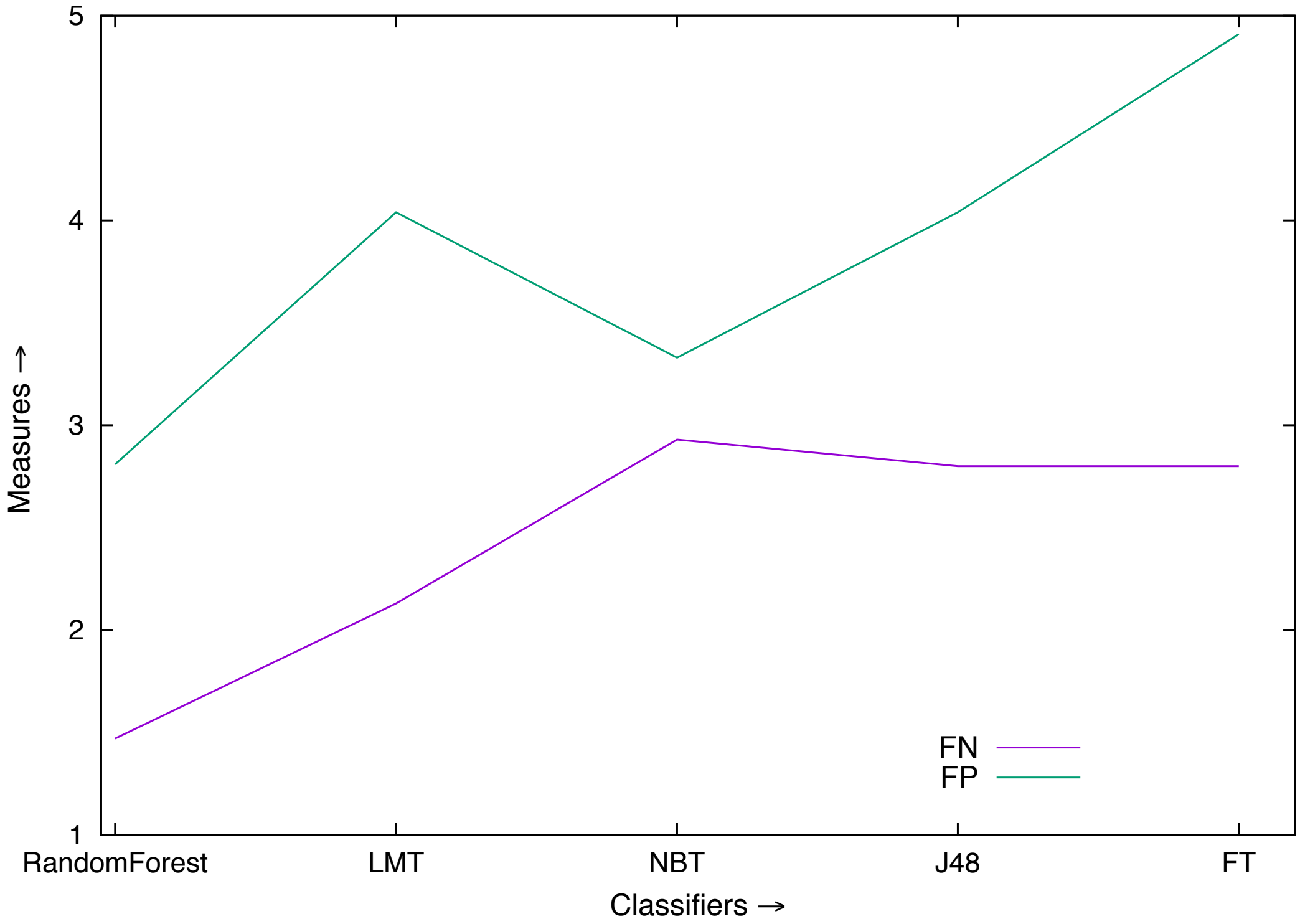

**Figure 11.** ***FP and FN of top five classifiers.***

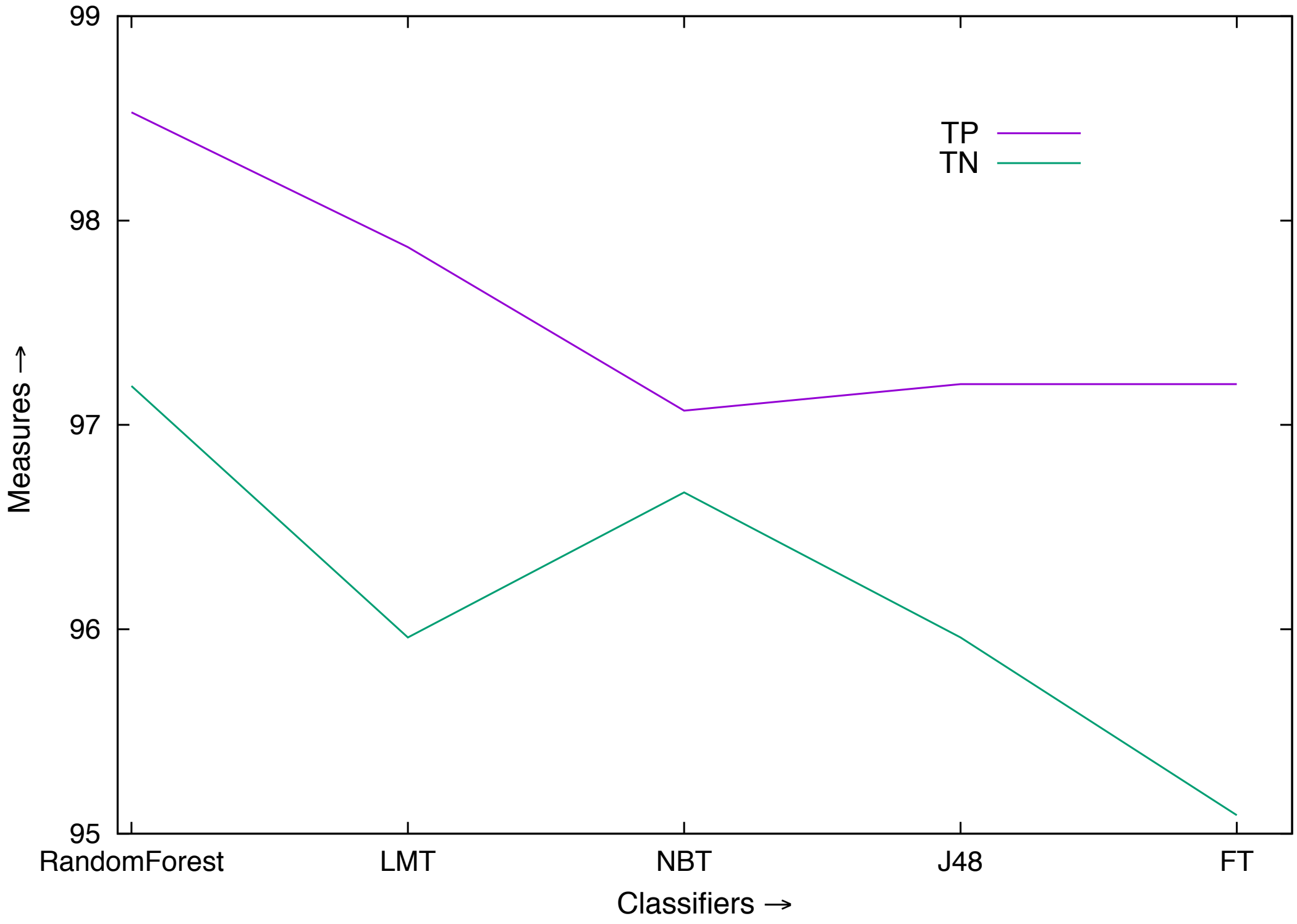

**Figure 12.** ***TP and TN of top five classifiers.***

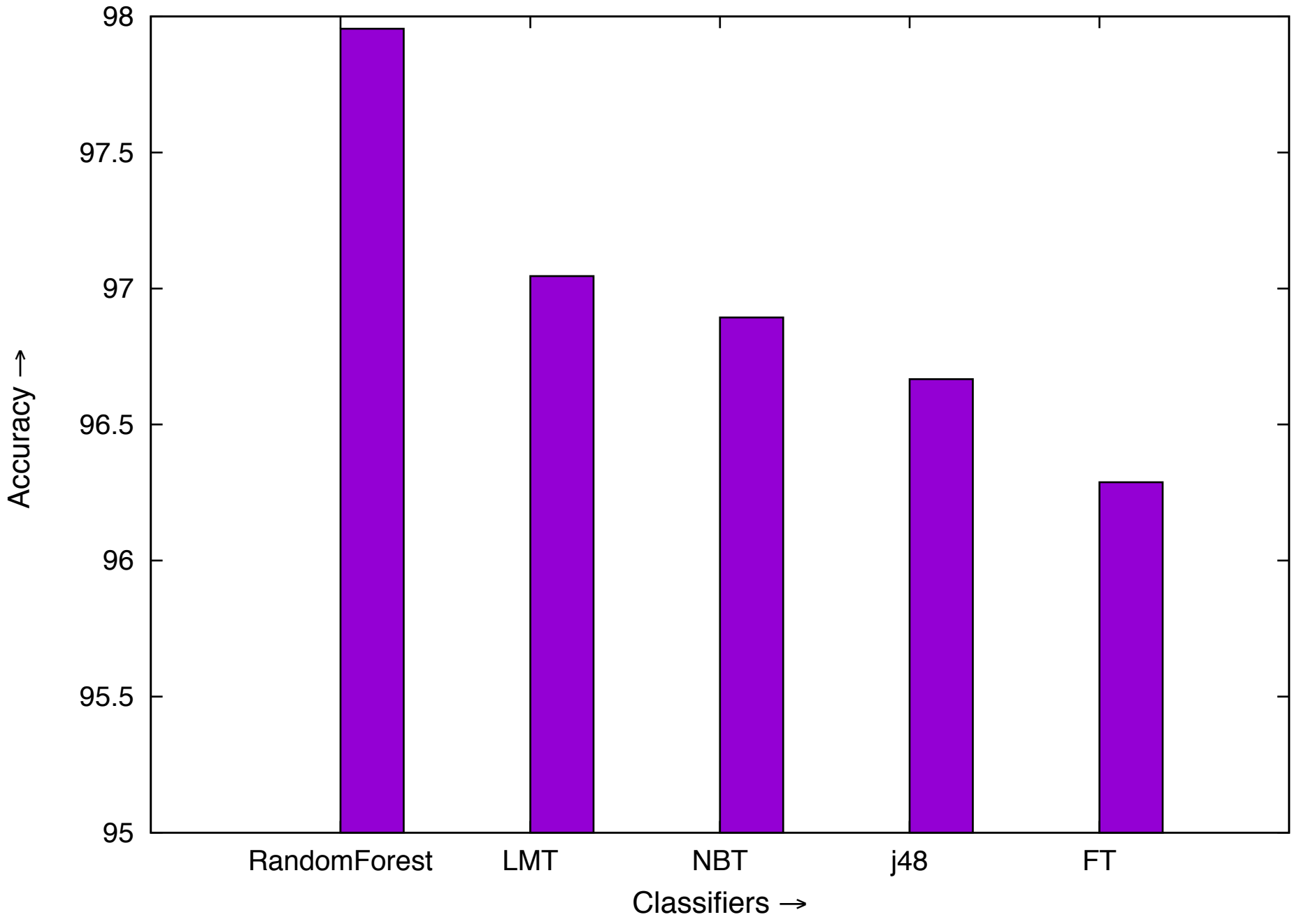

**Figure 13.** ***Accuracy of the top five classifiers.***

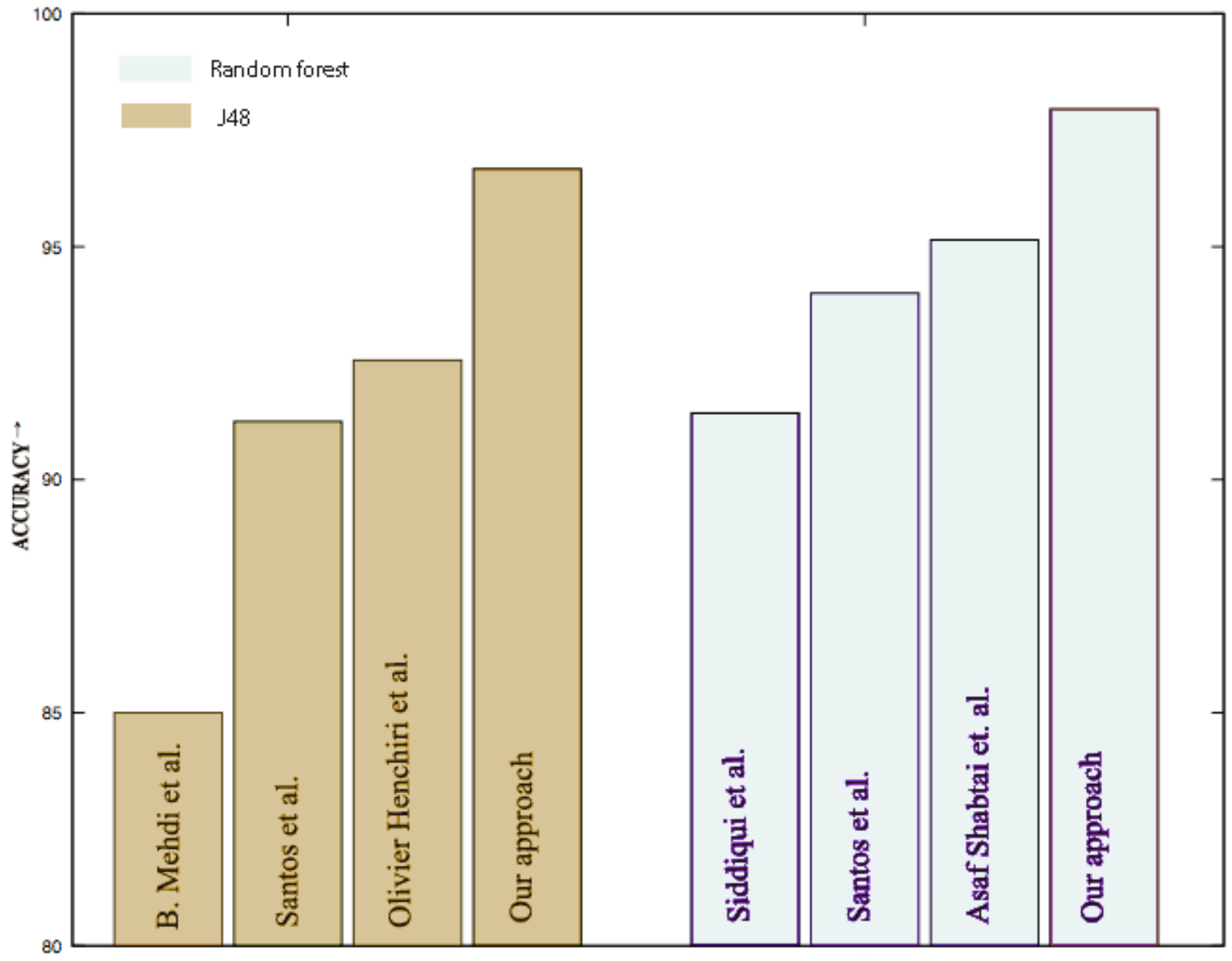

**Figure 14.** ***Comparison of the accuracy obtained by our approach and others.***

## 5. Conclusion

Traditional approach i.e. updating the signature database to combat advanced malware is ineffective. Therefore, in this paper, we presented a novel approach to detect the advanced malware with high accuracy. For the classification, we obtained the promising features (opcodes) by grouping the executables in 5 KB size. Extensive experiment has been done to study the performance of the classifiers viz. Random forest, LMT, NBT, J48 and FT in terms of TPR, TNR, FPR, FNR and accuracy by analyzing 11688 malware downloaded from malicia-project and 4006 benign programs collected from different systems. By our approach all five classifiers are able to uncover unknown malware with greater than 96.28% accuracy, which is better than the detection accuracy (~95.9%) reported by Santos et. al. (2013). Among these classifiers, we found that Random forest is the best (～97.95%) classifier to detect the unknown malware. Thus, our approach outperforms to detect the unknown malware and hence can be an effective technique to complement the signature based mechanism or dynamic analysis. In future, we will collect more malware and benign and will perform in-depth size analysis for the classification of unknown malware.

`

## Acknowledgments

Mr. Ashu Sharma is thankful to BITS, Pilani, K.K. Birla Goa Campus for the support to carry out his work through Ph.D. scholarship No. Ph603226/Jul. 2012/01. We are also thankful to IUCAA, Pune for providing hospitality and computation facility where part of the work was carried out.

## Authors

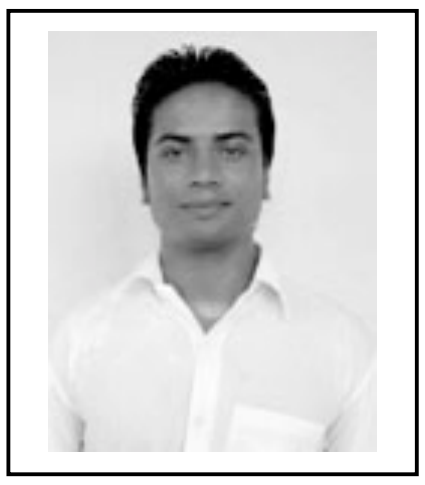

**Ashu Sharma** is a full time research scholar in Department of Computer Science and Information Systems, BITS-Pilani, K.K. Birla Goa Campus, India and perusing for his Ph.D degree on malware analysis under the supervision of Dr. Sanjay K. Sahay. He has published couple of papers in malware analysis and two papers are published in conference proceedings and will appear in Springer Verlag and Elsevier Procedia CS.

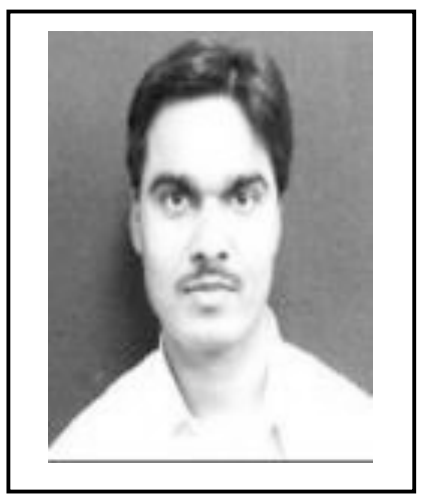

**Sanjay K. Sahay** received his Ph.D in 2003 and currently working as an Assistant Professor in Computer Science and Information System at BITS-Pilani, K.K. Birla Goa Campus, India. His research interest includes malware analysis, data mining, gravitational waves and machine learning. He has published papers on topics like malware analysis, gravitational waves, machine learning and data mining.